\documentclass[a4paper,11pt]{article}
\usepackage{amsfonts}
\usepackage{amsmath,amssymb,amscd}
\usepackage{appendix,marginnote,tikz,pgf,mathtools}
\usepackage{comment}  

\usepackage[numbers]{natbib}
\def\cL{\mathcal{L}}
\def\cM{\mathcal{M}}

\def\ll{\left\langle}
\def\rr{\right\rangle}

\def\be{\begin{equation}}
\def\ee{\end{equation}}
\def\ben{\begin{equation*}}
\def\een{\end{equation*}}
\def\bea#1\ena{\begin{align}#1\end{align}}
\def\bean#1\enan{\begin{align*}#1\end{align*}}

\def\pd{\partial}
\def\d{\partial}

\makeatletter

    \@addtoreset{equation}{section}
\makeatother

\baselineskip=\normalbaselineskip

\begin{document}

\begin{center}

{\Large \bf  
 Open KdV hierarchy of 2d minimal gravity of  Lee-Yang series
}
\vskip 1cm
{\large 
Hisayoshi Muraki\footnote{hmuraki@sogang.ac.kr}
 and Chaiho Rim\footnote{rimpine@sogang.ac.kr}
}\\
{Department of Physics, 
Sogang University, Seoul 04107, Korea}
 \end{center}
\vskip 10mm

\begin{abstract}
We present the open KdV hierarchy of 2d  minimal gravity of Lee-Yang series
which uses the  boundary cosmological constant as a flow parameter. 
The boundary cosmological constant is a conjugate variable to the boundary flow parameter 
used in the open KdV hierarchy of the intersection numbers on the moduli space of Riemann surfaces with boundaries.
The two generating functions are related through the Laplace transform.  
\end{abstract}

\tableofcontents

\section{Introduction} 

Minimal gravity is  a 2-dimensional quantum gravity coupled with minimal conformal matter,
so that the resulting theory still remains conformal and topological $(c=0)$ \cite{Kostov, Kazakov, KPZ}, 
and its correlation numbers obey the Gelfand-Dickey hierarchy in general \cite{BDSS, Doug, Ginsparg}. 
The minimal gravity of the Lee-Yang series is described by one-matrix model, 
whose correlation numbers show the KdV hierarchy, and thus is 
closely related with the intersection theory on the moduli space of Riemann surfaces.
In fact, by Witten, the generating function of the intersection theory 
had also been conjectured to satisfy the KdV hierarchy together with the string equation \cite{Witten}. 
Witten's conjecture for intersection theory was proved by Kontsevich using one-matrix model \cite{Konts}.
The generating functions for minimal gravity of Lee-Yang series
on $g=0,1,2$ have also been constructed using the KdV hierarchy \cite{BT, BBT}, and the resulting correlations 
(but off-shell, i.e., with arbitrary $t_k$ parameters but $t_1 \to t_1 +1$) have been shown to obey 
the recursion relations of topological gravity, as suggested by Witten.

Recently, the intersection theory is extended
to the moduli space of Riemann surfaces with boundaries \cite{PST,Buryak}. 
This extension is based on an extension of the KdV structure,
where the KdV hierarchy for closed surfaces has an equivalent description by the Virasoro constraints as shown in \cite{DVV, MMM}.
With boundaries, the generating function is described by the so-called open KdV hierarchy 
which contains the KdV flow parameters as well as a new boundary flow parameter $s$ 
indicating the boundary contribution.

It is naturally expected that the open KdV hierarchy also describes minimal gravity of Lee-Yang series on a disk. 
In our previous paper \cite{bmr_2018}, 
we check this expectation using the free energy on a disk,
where the boundary cosmological constant $\mu_B$ is introduced as an additional boundary parameter
that has a similar role of the boundary 
flow parameter $s$ in the intersection theory \cite{Tess, Ale2,Ale,Buryak2}.

The parameter $s$ looks natural in the the intersection theory. 
However, from the viewpoint of the 2d minimal gravity,
the boundary cosmological constant $\mu_B$ appears rather naturally and it is desired to have the open KdV hierarchy
in terms of $\mu_B$.
In this paper, we present the open KdV hierarchy using the boundary cosmological constant  $\mu_B$.

In section 2, we summarize the KdV hierarchy 
of the minimal gravity of the Lee-Yang series 
on the closed Riemann surfaces.
It is confirmed that the free energy known in the minimal gravity 
satisfies the KdV hierarchy along with the string equation.

In section 3, we investigate the KdV hierarchy on Riemann surfaces with boundaries. 
We use the known free energy of the minimal gravity on a disk 
to check the open KdV hierarchy and the open string equation.
The free energy is a function of the boundary cosmological 
constant $\mu_B$ rather than the $s$-parameter.
Therefore, one needs to find the relation of the hierarchy 
of the intersection theory with that of the minimal gravity. 
It turns out that $s$ and  $\mu_B$ are conjugate to each other.
We present the Virasoro constraints of the $\mu_B$-representation 
as their Laplace transforms of the $s$-representation and check those equivalence.
The open hierarchy allows one to find 
the free energy with genus $g\geq1$
using the free energy with genus 0 (disk). We demonstrate it for $g=1$.

Section 4 is devoted to the summary and discussion.
In appendices, useful identities used in the text can be found.

\section{ Lee-Yang series on a closed  Riemann surface}

Minimal gravity  of Lee-Yang series $\cM(2, 2p+1)$  is described by one-matrix model.
At the continuum limit, the matrix variable
is described by a differential operator   $\hat{Q}_2= \pd_x^2+u(x)$ whose dispersionless limit  (neglecting 
derivatives) reduces to a second order polynomial $Q_2= y^2+u(x)$. 
The polynomial $Q_2$ defines a one-dimensional Frobenius manifold $A_1$, $u$ being its coordinate.
This section describes how $u$ behaves according to
the flow equation on a Riemann surface without boundaries.

\subsection{Lee-Yang series on a sphere}

The Frobenius manifold allows flat coordinates 
and  the one-dimensional coordinate $u$ is trivial 
and obviously regarded as the flat one,
which will be renamed as $v$. 
The coordinate is a function of $x$. 
The variable $x$ can be identified with 
one of the variable of the generating function of the minimal gravity through the Douglas equation,
which obtains from the least action principle.
The string action describing the minimal gravity of Lee-Yang series $\cM(2,2p+1)$ on a sphere  is given by
\be
	S_{2/(2p+1)}= \theta_{1,p+1} +\sum_{n=0}^{p-1} t_n \theta_{1,n},
\ee 
where $\theta_{1,n}$ is defined by
\be	
	\theta_{1,n}
	=-\frac{\Gamma(1/2)}{\Gamma(1/2+n+1)}\underset{y=\infty}{\text{Res}}\left(Q_2^{1/2+n}\right)
	=\frac{v^{n+1}}{(n+1)!}.
\ee
The parameters $t_n$ are called the KdV parameters with the gravitational scaling dimensions (gcd)
(defined as the power of the cosmological constant $\mu$):
\be
	[t_n]=\frac{p+1-n}{2}, \qquad 0\le n \le p-1,
\ee 
where $t_0$ has the highest gcd.
It is noted that  $t_{p-1}$ has a special role
in the minimal gravity since  $[t_{p-1}]=1$
and is identified as the cosmological constant $\mu$.  
It is also useful to remember $[Q_2]=1/2$
so that $[v]=1/2$ and $[y]=1/4]$ in the Lee-Yang series.

The action principle results in the string equation:
Its derivative with respect to $v$ vanishes:
\begin{align}
\label{string_eq}
0&=\frac{\pd S_{2/(2p+1)}}{\pd v}=t_0 - x,
\\
\label{def-x}
	x&= -  \sum_{n=1}^{p+1}t_n\frac{v^n}{n!},
\end{align}
where we use the convention
 $t_{p+1}=1$ and $t_{p}=0$. 
The defining relation of $x$ in \eqref{def-x}
shows that $v$ is the function of $x$ and 
$t_n$: $v=v(x, \{t_{n \ne 0} \})$.
The variable $x$ has the same gcd as that of $t_0$
and is identified with $t_0$ 
if one applies the string equation \eqref{string_eq}.

To distinguish the $A_1$ coordinate $v$ 
from   $v( \{t_n \})$, 
the KdV parameter dependent variable,
we will use separate notations:
$v$ for the original $A_1$ coordinate  
and  $w $ for $v(x, \{t_{n \ne 0} \})$.
Then, $w$ satisfies the flow equations, 
consistent with the KdV hierarchy on a sphere:
\be\label{hierarch-sphere}
	\frac{\pd w}{\pd t_n}=\frac{\pd}{\pd x}\left(\frac{\pd \theta_{1,n+1}}{\pd w}\right) 
=\frac{w^n}{n!}\frac{\pd w}{\pd x}, \qquad  n \ge 1.
\ee

The free energy on a sphere $F_{\rm sphere}$
is known to have the form \cite{BZ}
\be \label{eq1}
	F_{\rm sphere}(\{t_m\}) =\frac{1}{2}\int_0^{w_*} 
d v\left[\mathcal{P}(v)\right]^2;
	\qquad \mathcal{P}(v) =\sum_{n=0}^{p+1} t_n  \frac{v^{n}}{n!}.
\ee 
$w_*$ is one of the solutions of the string equation 
\be\label{string-sphere-LY}
\mathcal{P}(w_*) =0,
\ee
so that 
$w_* = w (x=t_0, \{t_{n \ne 0} \} )$ which reduces to $\sqrt{-t_{p-1}}$ 
when $t_{n \ne (p-1, p+1)} \to 0$.
Since the only difference between 
$w_*$ and $w$ is the parameter $t_0$ or $x$,
we will omit the star if there is no confusion. 

The free energy \eqref{eq1} shows that the two-point correlation 
$\pd^2F_{\rm sphere}/{\pd t^2_0} =w $.
Multi-correlation is given in terms the derivatives of $F_{\rm sphere}$ with respect to $\{ t_m\}$:
\be
	  \frac{\pd^n F_{\rm sphere}}{\pd t_{a_1}  \cdots  \pd t_{a_n}}=\ll \prod_i^n O_{a_i} \rr_{{\rm sphere}}.
\ee
Especially, two-point correlation\footnote{This result is equivalent to the Liouville minimal gravity if the resonance transformation is considered \cite{BZ}.} 
is given as 
\be
\label{two-point-sphere}
 \ll O_0 O_{n-1} \rr_{{\rm sphere}} = \frac{w^{n}}{n!} .
\ee
One may evaluate the correlation on-shell  
if $t_n $ is set to be 0 for all $n$, except $t_{p-1}$ and $t_{p+1}=1$.

\subsection{Hierarchy on a closed Riemann surface}

Beyond the sphere one may generalize the KdV hierarchy
appearing in the intersection numbers  \cite{Witten}
\be\label{hierarchy-LY}
\frac{1}{\lambda^{2}}\frac{2n+1}{2}\frac{\pd^3 F^c}{\pd t^2_0\pd t_n}
=\frac{\pd^2 F^c}{\pd t^2_0}\frac{\pd^3 F^c}{\pd t^2_0\pd t_{n-1}}
+\frac{1}{2}\frac{\pd^3 F^c}{\pd t^3_0}\frac{\pd^2 F^c}{\pd t_0\pd t_{n-1}}
+\frac{1}{8}\frac{\pd^5 F^c}{\pd t^4_0\pd t_{n-1}} ,
\qquad n\geq 1,
\ee
with the string equation
\be
\label{string-closed}
\frac{\pd F^c}{\pd t_0}=\sum_{n\geq0}t_{n+1}\frac{\pd F^c}{\pd t_n}+\frac{t^2_0}{2\lambda^2}.
\ee

Note that the free energy $F^c$ has the genus expansion
\be
\label{genus-expansion-closed}
F^c=\sum_{g=0}^\infty \lambda^{2g-2}F^c_{(g)},
\ee
where  $\lambda$ is
a formal expansion parameter
and {$F^c_{(0)}=F_{\rm sphere}$}.
If one has the genus 0 part of 
\eqref{hierarchy-LY} and \eqref{string-closed},
one compares with the KdV hierarchy 
\eqref{hierarch-sphere} and 
the string equation \eqref{string-sphere-LY}
in the Lee-Yang series,
one finds $t_1$ is to be shifted by 1 ($t_1 \to t_1+1$) \cite{BT,BBT}.
Therefore, 
the hierarchy on a closed Riemann surface 
has the form \eqref{hierarchy-LY}
but 
the string equation for the Lee-Yang series
has to be modified 
\be
\label{string-closed-LY}
0=\sum_{n\geq0}t_{n+1}\frac{\pd F^c}{\pd t_n}+\frac{t^2_0}{2\lambda^2}.
\ee

\subsection{Lee-Yang series on a torus}

The  hierarchy and  string equation 
have the form at genus 1:
\begin{align}
\label{KdV-torus1}
 \frac{2n+1}{2}\frac{\pd^3 F^c_{(1)}}{\pd t^2_0\pd t_n}
& =\frac{\pd^2 F^c_{(0)}}{\pd t^2_0}\frac{\pd^3 F^c_{(1)}}{\pd t^2_0\pd t_{n-1}}
+\frac{1}{2}\frac{\pd^3 F^c_{(0)}}{\pd t^3_0}\frac{\pd^2 F^c_{(1)}}{\pd t_0\pd t_{n-1}}
\nonumber\\
&\quad\quad
+\frac{\pd^2 F^c_{(1)}}{\pd t^2_0}\frac{\pd^3 F^c_{(0)}}{\pd t^2_0\pd t_{n-1}}
+\frac{1}{2}\frac{\pd^3 F^c_{(1)}}{\pd t^3_0}\frac{\pd^2 F^c_{(0)}}{\pd t_0\pd t_{n-1}}
+\frac{1}{8}\frac{\pd^5 F^c_{(0)}}{\pd t^4_0\pd t_{n-1}},
\\
\label{string-torus1}
0&= \sum_{n\geq0}t_{n+1} \frac{\pd F^c_{(1)}}{\pd t_n}.
\end{align}
One may simplify \eqref{KdV-torus1}
if one multiplies \eqref{KdV-torus1} with $t_n$
and sums over $n\ge 1$:
\be
\sum_{n\ge 0} 
 \frac{2n+1}{2} t_n \frac{\partial^3  F^c_{(1)}}
 {\partial t_0^2 \partial  t_n}
=-\frac{\partial^2  F^c_{(1)}} {\partial t_0^2 },
\ee
or
\be
\label{KdV-torus1-reduced}
\frac{\partial^2}{\partial t_0^2}
\left(
\sum_{n\ge 0}  
 \frac{2n+1}{2} t_n \frac{\partial  F^c_{(1)}}
 { \partial  t_n}  \right)=0.
\ee 
Here we use the string equation at genus 0  \eqref{string-closed-LY} and \eqref{string-torus1}.

The free energy on a torus is known in \cite{BT,BBT}
\be\label{eq:Fc:1}
F^c_{(1)}=a  \log{P'(w) }.
\ee
 Here $P'(w) $ stands for $\partial P(w)/\partial w$ with $t_n$'s fixed.  
The coefficient $a$ is a constant and is found $ a = -1/{24}$ (see appendix B for detailed derivations\footnote{
In \cite{BT,BBT}, the notations for free energies $F^c_{(g)}$ in matrix model description is normalized so that the free energy is to be rescaled by a factor of $2^g$.}).
One can check that  the free energy \eqref{eq:Fc:1}
satisfies \eqref{string-torus1} and 
 \eqref{KdV-torus1-reduced}.
 The proof goes as follows. 
 Note that  $P'(w)$ is the function of 
 $w_n$'s and $t_n$'s:
\be
\label{P'}
P'=\sum_{n\ge1} t_n \frac{w^{n-1}}{ (n-1)!},
\ee 
where $w$ is again the function of $t_n$'s
through the KdV equation. 
For example,
\be
0 =\frac{\pd P'(w, t)}{\pd t_n}
 =  \left( \frac{\partial P'(w, t)}{\partial  t_n } \right)_{w}
 + \left( \frac{\partial P'(w, t)}{\partial  w } \right)_{t} \frac{\partial  w} {\partial t_n},
\ee
where $()_a$ means differentiation with $a$ fixed.

One can check the  string equation \eqref{string-torus1}, using the relation
\be 
\sum_{n\ge 0} t_{n+1} \frac{\pd P'}{\pd t_n}
=\sum_{n\ge 1} t_{n+1} \frac{w^{n-1}}{(n-1)!}
+ P'' \sum_{n\ge 0} t_{n+1} 
\frac {\pd w}{\pd t_n} =0,
\ee
where we used the second derivative of \eqref{string-closed-LY} and an identity $P''= \sum_{n\ge 1} t_{n+1} \frac{w^{n-1}}{(n-1)!}$.

The simplified hierarchy \eqref{KdV-torus1-reduced} is evaluated as follows: Note that
 \be  
\sum_{n\ge 0}  
 \frac{2n+1}{2} t_n \frac{\partial  P'}
 { \partial  t_n}=  
\sum_{n\ge 0}  
 \frac{2n+1}{2} t_n\left(\frac {\partial  P'}{\partial t_n } \right)_{w}
 +
  w P'   P''  \,\frac{\partial w}{\partial t_0},
\ee 
using \eqref{string-sphere-LY} and \eqref{P'}.
The first term of the right hand side is written as 
\be
\sum_{n\ge 0}  
 \frac{2n+1}{2} t_n\left(\frac {\partial  P'}{\partial t_n } \right)_{w} = (wP')' + \frac{P'}{2},
\ee
if one uses \eqref{P'} and $(w P')' = \sum_{n\ge0 } n t_n \frac{w^{n-1}}{ (n-1)!}$.
In addition, we have the null equality 
\begin{align} 
\label{eq:derivative:of:P}
0 =\frac{\pd P(w, t)}{\pd t_0} = \left( \frac{\partial P(w, t)}{\partial  w } \right)_{t}
   \frac{\partial  w} {\partial t_0}
 +  \left( \frac{\partial P(w, t)}{\partial  t_0 } \right)_{w}
 =P' \frac{\partial  w} {\partial t_0}  + 1.
\end{align}
This shows that
 \be  
\sum_{n\ge 0}  
 \frac{2n+1}{2} \frac{t_n}a  \frac{\partial  F^c_{(1)} }
 { \partial  t_n}= 
 \frac{(wP')'}  {P'} + \frac {1}2 
 - w  \frac{ P''}{P'}   
= \frac {3}2,
\ee 
which satisfies \eqref{KdV-torus1-reduced}.


\section{Open KdV hierarchy of minimal gravity with boundaries}

\subsection{Open KdV in the intersection theory}
A similar KdV hierarchy (``open KdV hierarchy") has been proposed for intersection theory 
on the moduli space of Riemann surfaces with boundaries, using an additional flow parameter $s$.
The flow along $t_n$ is given as  \cite{PST}
\be\label{hierarchy-open}
	\frac{2n+1}{2}\frac{\pd F^o}{\pd t_n}
	=\lambda\frac{\pd F^o}{\pd s}\frac{\pd F^o}{\pd t_{n-1}}
	+\lambda\frac{\pd^2 F^o}{\pd s\pd t_{n-1}}
	+\frac{\lambda^2}{2}\frac{\pd F^o}{\pd t_0}\frac{\pd^2 F^c}{\pd t_0\pd t_{n-1}}
	-\frac{\lambda^2}{4}\frac{\pd^3 F^c}{\pd t^2_0\pd t_{n-1}}\,,
	\qquad n\geq1.
\ee
The  open string equation  is given by
\be
\label{string-open}
	\frac{\pd F^o}{\pd t_0}=\sum_{n\geq0}t_{n+1}\frac{\pd F^o}{\pd t_n}+\frac{s}{\lambda}.
\ee
In addition, the open KdV together with the string equation is shown to be equivalent to 
the Virasoro constraints with additional flow equation of $s$  \cite{Buryak}
\be
\label{s-flow}
\frac{\pd F^o}{\pd s}
= \lambda \left(
\frac12 \left( \frac{\pd F^o}{\pd t_0}\right)^2
+\frac12 \frac{\pd^2  F^o}{\pd t_0^2}
+\frac{\pd^2  F^c}{\pd t_0^2}
\right).
\ee

The free energy is expanded  
in the genus expansion
\be\label{genus-expansion-open}
	F^o=\sum_{g=0}^\infty \lambda^{g-1}F^o_{(g)},
\ee
whose lowest order  ($g=0$) gives
\bea
\label{g=0-o-kdv}
	\frac {2n+1}2  \frac{\pd F^o_{(0)}}{\pd t_n}
	&= 
	\frac{\pd  F^o_{(0)}}{\pd s}\frac{\pd F^o_{(0)}}{\pd t_{n-1}}+
	\frac{1}{2}\frac{\pd F^o_{(0)}}{\pd t_0}\frac{\pd^2 F^c_{(0)}}{\pd t_0\pd t_{n-1}} ,
	\\
\label{g=0-string}
	0&=\sum_{n\geq0}t_{n+1}\frac{\pd F^o_{(0)} }{\pd t_n}+s ,
	\\
\label{g=0-s-flow}
	\frac{\pd F^o_{(0)}}{\pd s}
	&= 
	\frac12  \left(\frac{\pd F^o_{0}}{\pd t_0} \right)^2 +\frac{\pd^2  F^c_{(0)} } {\pd t_0^2} .
\ena

\subsection{Free energy on a disk and hierarchy}

One may conjecture that the free energy  $F_{\rm disk}$ on a disk \cite{MSS, IR}
can be a  solution  of the hierarchy \eqref{g=0-o-kdv}  \cite{bmr_2018}.
It is noted that  the free energy  $F_{\rm disk}$
corresponds to the continuum limit of  the trace of $\log (M + \mu_B)$ 
where the one-matrix element is replaced by $Q$
and $\mu_B$ is the boundary cosmological constant.
The integral representation is given as follows:
\begin{align}\label{Gen.Funct.Disk}
	F_{\rm disk}
	 &=\frac{i}{\sqrt{2\pi}}\int_0^\infty \frac{dl}{l} \ e^{-l\mu_B}
		\int_{t_0}^\infty dx \int_{\mathbb{R}} dy \ e^{ -l ( y^2+w) }.
\end{align}
Here we assume a proper regularization (subtracting the infinity)
in the integration limit at $l \to 0$.
In fact, $F_{\rm disk}$ satisfies a similar equation as in  \eqref{g=0-o-kdv}
\be
	\label{g=0-o-kdv-FB}
	\frac {2n+1}2  \frac{\pd F_{\rm disk}}{\pd t_n}
	= 
	-\mu_B \frac{\pd F_{\rm disk}}{\pd t_{n-1}}+
	\frac{1}{2}\frac{\pd F_{\rm disk}}{\pd t_0}\frac{\pd^2 F^c_{(0)}}{\pd t_0\pd t_{n-1}}.
\ee
The identification goes along with 
the $s$-flow relation  \eqref{g=0-s-flow} \cite{bmr_2018}:
\be 
\label{g=0-s-flow-FB}
-\mu_B
= \frac12  
\left(\frac{\pd F_{\rm disk}} {\pd t_0} \right)^2 
+\frac{\pd^2  F^c_{(0)} } {\pd t_0^2} .
\ee

\subsection{Open hierarchy with $\mu_B$}

The open KdV formula in \eqref{g=0-o-kdv-FB}
 obtained using the free energy on a disk 
is to be compared with  the one   in  \eqref{g=0-o-kdv}
from the intersection theory.
The result is 
\be
\label{s-muB-1-FB}
 	{\frac{\pd F^o_{(0)}}{\pd s}} = -\mu_B.
\ee
On the other hand, if  the string equation
\eqref{g=0-string} is used, 
the parameter $s$   turns out 
\be
\label{s-muB-2-FB}
s= \frac{\partial F_{\rm disk}(\mu_B)}{\partial \mu_B}, 
\ee   
which is identified as the loop operator 
which is the continuum limit of the resolvent,  trace of $ 1/(M + \mu_B)$.
This identification raises a problem: 
$s$ identified in \eqref{s-muB-2-FB}
depends on the KdV variables $\{t_n\}$
and spoils the property of $s$ 
that $s$ should  be 
independent of $\{t_n\}$
as conjectured in the intersection theory.

To cure this problem, one notes that $s$ and $\mu_B$ are conjugate according to 
\eqref{s-muB-1-FB} and \eqref{s-muB-2-FB}. 
Therefore, one may use either $s$ or $\mu_B$, not both. 
According to the free energy of minimal gravity,
it is desirable to put the hierarchy in terms of $\mu_B$ as given in 
\eqref{g=0-o-kdv-FB}.
One may follow the same steps 
in \cite{Buryak} using the half Burger-KdV hierarchy
except the two changes:
One is ${\pd F^o_{(0)}}/{\pd s} \to  -\mu_B$
using the fact \eqref{s-muB-1-FB}
and the other is  
$ s \to {\partial F_{\rm disk}(\mu_B)}/{\partial \mu_B}$ 
as noted in \eqref{s-muB-2-FB}. 
Then one has $\mu_B$-representation.

The open KdV hierarchy on a disk is given as 
\eqref{g=0-o-kdv}.
The relation of the boundary parameter is given 
as \eqref{g=0-s-flow-FB}.
Finally, the open string equation (after  $t_1 \to t_1+1$) 
is modified as 
\be	
\label{g=0-string-FB}
0=\sum_{n\geq0}t_{n+1}\frac{\pd F_{\rm disk} }{\pd t_n}+\frac{\pd F_{\rm disk} }{\pd \mu_B}.
\ee
According to this $\mu_B$-representation,
one may have  the open KdV hierarchy  
\be
\label{open-kdv-muB}
	\frac{2n+1}{2}\frac{\pd F^o}{\pd t_n}
	=- \mu_B \frac{\pd F^o}{\pd t_{n-1}}
	+\frac{\lambda^2}{2}\frac{\pd F^o}{\pd t_0}\frac{\pd^2 F^c}{\pd t_0\pd t_{n-1}}
	-\frac{\lambda^2}{4}\frac{\pd^3 F^c}{\pd t^2_0\pd t_{n-1}},	\qquad n\geq1,
\ee
the open string equation 
\be\label{open-string-muB}
	0=\sum_{n\geq0}t_{n+1}\frac{\pd F^o}{\pd t_n}+\frac{\pd F^o}{\pd  \mu_B},
\ee 
and the boundary parameter constraint 
\be
\label{open-constraint-muB}
-\mu_B 
= \lambda^2 \left(
\frac12 \left( \frac{\pd F^o}{\pd t_0}\right)^2
+\frac12 \frac{\pd^2  F^o}{\pd t_0^2}
+\frac{\pd^2  F^c}{\pd t_0^2}
\right).
\ee

\subsection{Virasoro constraints}

The $\mu_B$-representation 
is obtained by replacing $s \to \pd /\pd \mu_B$ and $\pd/\pd s \to - \mu_B$.
The hierarchy is equivalent to put (the half set of) the Virasoro generators of form 
\be
\label{eq:Virasoro:gen:op}
{\cal L}_n = L_n + (-\mu_B)^n 
\left( -\mu_B\frac{\pd }{\pd \mu_B} 
- \frac{n+1}4  \right), \qquad n\geq -1,
\ee
which  satisfies the commutation relation
\be
\label{eq: commutation}
[\cL_n,\cL_m]=(n-m)\cL_{n+m}.
\ee

$L_n$ is the Virasoro generator of the closed surface\footnote{
We use the original $t_1$ before shifted by 1}:
\be
\begin{aligned}
\label{eq:Virasoro:gen:cl}
L_n:=\sum_{i\ge 0}\frac{(2i+2n+1)!!}{2^{n+1}(2i-1)!!}(t_i-\delta_{i,1})\frac{\d}{\d t_{i+n}}+\frac{u^2}{2}\sum_{i=0}^{n-1}\frac{(2i+1)!!(2n-2i-1)!!}{2^{n+1}}\frac{\d^2}{\d t_i\d t_{n-1-i}}\\
+\delta_{n,-1}\frac{t_0^2}{2 u^2}+\delta_{n,0}\frac{1}{16},
\end{aligned}
\ee
which imposes the Virasoro constraints on the tau function of intersection theory of closed surfaces
\be
\label{eq: Virasoro}
L_n\exp(F^c)=0, \qquad n\ge -1.
\ee

One may show that the partition function $\exp(F^o+F^c)$ 
is constrained by the Virasoro generator \eqref{eq:Virasoro:gen:op}:
\be
\label{eq: open virasoro}
\cL_n\exp(F^o+F^c)=0,\qquad n\ge -1.
\ee

The $\mu_B$-representation of $\cL_n$  is obtained 
if one uses the Laplace transformation from the original $s$-representation 
of the partition function
\be
\exp(F^o+F^c)(s) = \int d\mu_B \, e^{-s\mu_B} \exp(F^o+F^c)(\mu_B),
\ee
with an appropriate integration contour so that the integration converges.
The Laplace transform ensures the replacements: $s \to \pd /\pd \mu_B$ and $\pd/\pd s \to - \mu_B$.
Therefore, one may expect that the Virasoro constraint in $\mu_B$-representation
will result in the open analog of the Virasoro constraints \eqref{eq: Virasoro}.
The proof can be carried out in the $\mu_B$-representation 
directly in a parallel manner with that of the original open KdV hierarchy \eqref{hierarchy-open} presented in \cite{Buryak}.
For $n=-1$, the equation is equivalent to the open string equation \eqref{open-string-muB}, 
so that it is true by assumption.
Then it is sufficient to prove the open Virasoro constraints \eqref{eq: open virasoro} only for $n=0,1,2$,
because it follows from the commutation relation \eqref{eq: commutation} that $\cL_n=\frac{(-1)^{n-2}}{(n-2)!}ad_{\cL_1}^{n-2}\cL_2$, for $n\ge 3$.
The proof relies on an inductive relation (direct computations are given in appendix C)
\be
\label{eq:proof:induc}
	\frac{\cL_n\tau}{\tau}+\mu_B\frac{\cL_{n-1}\tau}{\tau}=0,
	\qquad  n=0,1,2,
\ee
where $\tau=\exp(F^o+F^c)$. From \eqref{open-string-muB} and \eqref{eq:proof:induc} it follows that 
$\cL_0\tau=-\tau\mu_B\frac{\cL_{-1}\tau}{\tau}=0$ (from the open string equation), 
and, by successive use of inductive relation \eqref{eq:proof:induc}, 
one has $\cL_1\tau=0$, and then $\cL_2\tau=0$.

\subsection{Free energy on a cylinder}

In this section, we present a solution to the open hierarchy with genus 1.
The genus expansions of \eqref{open-kdv-muB}, \eqref{open-string-muB} and  \eqref{open-constraint-muB}, respectively, have the forms
at each order of $g$:
\bea
	\frac{2n+1}{2}\frac{\pd F^o_{(g)}}{\pd t_n}
	&=-\mu_B \frac{\pd F^o_{(g)}}{\pd t_{n-1}}
		+	
		\frac12 \sum_{g_1+2 g_2=g}
		\left( 
		\frac{\pd F^o_{(g_1)}}{\pd t_0}
		\frac{\pd^2 F^c_{(g_2)}}{\pd t_0\pd t_{n-1}}
		\right)
		-
		\frac14 \frac{\pd^3 F^c_{\left((g-1)/2\right)}}{\pd t^2_0\pd t_{n-1}},
		\qquad n\geq1,
	\\
	0&=\sum_{n\geq0}t_{n+1}\frac{\pd F^o_{(g)} }{\pd t_n}+\frac{\pd F^o_{(g)}}{\pd  \mu_B}, \qquad g\ge0,
	\\
	-\mu_B & = \frac12  \frac{\pd F^o_{(0)}}{\pd t_0} \frac{\pd F^o_{(0)}}{\pd t_0} +\frac{\pd^2  F^c_{(0)} } {\pd t_0^2} ,
	\\
	0&=\sum_{g_1 + g_2=g} \frac12  \frac{\pd F^o_{(g_1)}}{\pd t_0} \frac{\pd F^o_{(g_2)}}{\pd t_0} 
	+\frac12 \frac{\pd^2  F^o_{(g-1)}}{\pd t_0^2}+\frac{\pd^2  F^c_{\left( g/2 \right)} } {\pd t_0^2}, \quad g\ge1,
\ena
where, needless to say, the term involving $F^c_{(g'/2)}$ is absent when $ {g'}/{2}$ is not an integer.

The free energy on a disk satisfies 
the lowest order ($g=0$) hierarchy.
Therefore, we have
\be
F^o_{(0)}= F_{\rm disk}.
\ee
The next order ($g=1$) has the following equations:
the open KdV hierarchy 
\be
	\label{g=1-o-kDv}
	\frac {2n+1}2  \frac{\pd F^o_{(1)}}{\pd t_n}
	= 
	-\mu_B \frac{\pd F^o_{(1)}}{\pd t_{n-1}}+
	\frac{1}{2}\frac{\pd F^o_{(1)}}{\pd t_0}\frac{\pd^2 F^c_{(0)}}{\pd t_0\pd t_{n-1}} 
- \frac14 \frac{\pd^3 F^c_{(0)}}{\pd t^2_0\pd t_{n-1}}, 
\ee
the string equation 
\be
\label{g=1-o-string}
0=\sum_{n\geq0}t_{n+1}\frac{\pd F^o_{(1)} }{\pd t_n}+\frac{\pd F^o_{(1)}}{\pd  \mu_B},
\ee
and the constraint equation for the boundary parameter
\be 
\label{g=1-o-constraint}
0= 
  \frac{\pd F^o_{(0)}}{\pd t_0}   \frac{\pd F^o_{(1)}}{\pd t_0}  
+\frac12 \frac{\pd^2  F^o_{(0)}}{\pd t_0^2}
= \frac{\pd F^o_{(0)}}{\pd t_0}   \frac{\pd}{\pd t_0}  
\left(  F^o_{(1)}+ \frac 12 \log  \left( \frac{\pd F^o_{(0)}}{\pd t_0} \right) 
\right).
\ee
The constraint equation \eqref{g=1-o-constraint} hints at the solution of the form
\be
\label{g=1-solution-o}
	 F^o_{(1)}=   - \frac 12 \log  \left( \frac{\pd F^o_{(0)}}{\pd t_0} \right) + f,
\ee 
where $f$ is a $t_0$-independent function.
One can easily check that $F^o_{(1)}$ and $F^o_{(1)}+cF^o_{(0)}$, for any constant $c$,
obey the same hierarchy \eqref{g=1-o-kDv} since $F^o_{(0)}$ 
satisfies the ($g=0$)-hierarchy \eqref{g=0-o-kdv-FB}.
Thus $f$ can be a function of $F^o_{(0)}$,
but which contradicts the fact that $f$ should be independent of $t_0$.
A natural choice is $f=0$.
Therefore, we conclude that the free energy at $g=1$ has the form
\be
\label{g=1-solution-open}
 F^o_{(1)}=   - \frac 12 \log  \left( \frac{\pd F^o_{(0)}}{\pd t_0}
\right) .
\ee 
One can check that $F^o_{(1)}$ in \eqref{g=1-solution-open} satisfies \eqref{g=1-o-kDv},
noting that \eqref{g=1-o-kDv} reduces to 
the $t_0$ derivative of \eqref{g=0-o-kdv-FB}:
\be
\frac{\pd}{\pd t_0} 
\left(\frac {2n+1}2  \frac{\pd F^o_{(0)}}{\pd t_n}+\mu_B \frac{\pd F^o_{(0)}}{\pd t_{n-1}}
-\frac{1}{2}\frac{\pd F^o_{(0)}}{\pd t_0}\frac{\pd^2 F^c_{(0)}}{\pd t_0\pd t_{n-1}}
\right) =0. 
\ee
Finally, it is obvious that $F^o_{(1)}$ satisfies the string equation 
\eqref{g=1-o-string} since $F^o_{(0)}$ satisfies the ($g=0$)-string equation \eqref{g=0-string-FB}.

\section{Summary and discussions}

We consider the open KdV hierarchy for 2d minimal gravity of Lee-Yang series. 
The hierarchy is given to have the boundary cosmological constant $\mu_B$ 
rather than the boundary flow parameter $s$ appearing in the intersection theory of open surfaces with boundaries. 
It is noted that $\mu_B$ and $s$ are conjugate variables and 
give rise to the Laplace transform of the free energy of 2d minimal gravity,
resulting in rephrasing the Virasoro constraint equations in terms of $\mu_B$.
It is noted, however, that the Laplace transform of the partition function 
does not need to produce the result of the intersection theory 
since the initial conditions of the two approaches are different.

The explicit form of the free energy at each order of genus $g$ 
can be obtained according to the closed and open KdV hierarchies
starting with the free energy of $g=0$. 
As an example, we present the free energy at $g=1$ in the text.

The Lee-Yang series $\cM(2,2p+1)$ is obtained from the one-matrix model.
The dual picture of the Lee-Yang series  (weak strong duality $b \leftrightarrow 1/b$ in the Liouville gravity) is described 
by the $A_{2p}$ Frobenius manifold \cite{BDM,BelavinRud}.
We expect the dual picture will show very different behavior and shall be worth studying. 
This is because the KdV hierarchy of the one-matrix model
in this dual picture does not work anymore and is to be replaced by the Gelfand-Dickey hierarchy.
In addition, the flow parameter of the dual theory 
coincides with the conformal parameter in the Liouville minimal gravity \cite{BelavinRud},
which is in contrast with the original $A_1$ description 
since the resonance transformation among those parameters
plays a central role  in general \cite{BZ, BMR,ABR}.
As a result, the open hierarchy in the dual picture 
will show very different behavior unlike the open KdV hierarchy.
We will provide this new feature in a separate paper.

\subsection*{Acknowledgements}
The work was partially supported by National Research Foundation of Korea  grant number \\ 2017R1A2A2A05001164.

\section*{Appendix A}

We provide a proof for the identity \eqref{g=0-o-kdv-FB} without explicit evaluation of the 
integral as in  \cite{bmr_2018}.
The proof goes as following. 
Taking derivatives of $F^o_{(0)}$ with respect to $t_n$ and using the KdV flow equation \eqref{hierarch-sphere},
one has the (off-shell) one-point correlation  
\begin{align}
	\ll O_n \rr_{\rm disk}
& 	=-\frac{i}{\sqrt{2\pi}}\int_0^\infty dl \ e^{-l\mu_B}
		\int_{t_0}^\infty dx \frac{\pd w}{\pd x}\frac{w^n}{n!} \int_{\mathbb{R}} dy   \ e^{ -l ( y^2+w) }
\nonumber\\
& =-\frac{i}{\sqrt{2\pi}}\int_0^\infty dl \ e^{-l\mu_B}
		\int_{w_*}^\infty dw \frac{w^n}{n!} \int_{\mathbb{R}} dy   \ e^{ -l ( y^2+w) },
\end{align}
where $w_*$ is the value of $w$ at $x=t_0$.
On the other hand,  the correlation multiplied by the boundary cosmological constant 
has the form if one uses the identity 
$-\mu_Be^{-l\mu_B} = {\partial e^{-l\mu_B}}/{\partial l}$
and integrates by part  with respect to $l$ 
\be 
	\mu_B\ll O_{n-1} \rr_{\rm disk} 
=\frac{i}{\sqrt{2\pi}}\int_0^\infty dl \ e^{-l\mu_B}
		\int_{w_*}^\infty dw  \frac{w^{n-1}}{(n-1)!} 
 \int_{\mathbb{R}} dy (y^2+w) \ e^{ -l ( y^2+w) } .
\ee
Here the boundary term at $l \to 0$ vanishes 
due to the proper regularization.

What we trying to prove is the  difference between 
$\ll O_n \rr_{\rm disk}$ and $\mu_B\ll O_{n-1} \rr_{\rm disk} $ .
This can be done using the following two identities.
One is the total derivative with respect to $y$ whose integrated  value is zero:
\bean
	0
	&=   \int_{w_*}^{\infty}dw 
\frac{ w^{n-1}}{ (n-1)!}
 \int_{\mathbb{R}}dy  \frac{d}{dy} 
		\left[\frac{ y}{l}  {e^{-l(y^2+w)} }  \right]\\
	&=  \int_{w_*}^{\infty}dw \int_{\mathbb{R}}dy \left(\frac{1}{l} - 2 y^2\right)\frac{w^{n-1}}{(n-1)!} \frac{\pd w}{\pd x}\,e^{-l(y^2+w)} .
\enan 
The other one is total derivative with respect to $w$:
\ben
	-  \int_{\mathbb{R}}dy  \left[ \frac{f(w)}{l}\,e^{-l(y^2+w)} \right]_{w=w_*}
	=  \int_{w_*}^{\infty}dw \int_{\mathbb{R}}dy  
	 \left[ \frac{1}{l}\frac{\pd f(w)}{\pd w}  -f(w)\right]e^{-l(y^2+w)},
\een
where $f(w)$ is an arbitrary polynomial. 
Then, one may find the linear relation with coefficients $c$ and $\alpha$  such that 
\be\label{integrand:of:rec:rel}
	-c \frac{w^n}{n!}   + \frac{w^{n-1}}{(n-1)!}(y^2+w) = \alpha \left(\frac{1}{l} - 2 y^2\right)\frac{w^{n-1}}{(n-1)!} + \frac{1}{l}\frac{\pd f(w)}{\pd u}  - f(w).
\ee
First, one eliminates the $(1/l)$-term in  \eqref{integrand:of:rec:rel}  so that 
$f(w)$ satisfies the condition
$ 	{\pd f(w)}/{\pd w} =- {\alpha w^{n-1}}/{(n-1)!}$ 
which fixes  
\be
	f(w)=-\alpha \frac{w^n}{n!}.
\ee
In addition, one may eliminate the $y$-dependent term in \eqref{integrand:of:rec:rel} by fixing $c$ and $\alpha$: 
\be
	\alpha=-\frac{1}{2}, \qquad c = \frac{2n+1}{2}.
\ee 
This results in the desired  the recursive relation \eqref{g=0-o-kdv-FB}:
\begin{align}
	0
		&
	=\frac{2n+1}{2} \ll O_n\rr_{\rm disk}  + \mu_B \ll O_{n-1}\rr_{\rm disk}
	-\frac{1}{2} \left[ \frac{w_*^n}{n!}\right] 
	\frac {i}{\sqrt{2\pi}}
	\int_0^\infty dl \  e^{-l\mu_B} \int_{\mathbb{R}}dy  \left[- \frac{e^{-l(y^2+w)}}{l} \right] 
\nonumber\\
	&
	=\frac{2n+1}{2} \ll O_n\rr_{\rm disk}  + \mu_B \ll O_{n-1}\rr_{\rm disk}
	-\frac{1}{2}  \ll O_0O_{n-1}\rr_{\rm sphere} \ll O_{0}\rr_{\rm disk},
\end{align}
where we use the result for two-point correlation on a sphere \eqref{two-point-sphere}.

\section*{Appendix B}

We may check that the coefficient $a$ in \eqref{eq:Fc:1} turns out to be $-1/24$.
As $a$ in \eqref{eq:Fc:1} is independent of 
$n$, here we demonstrate it using the case for $n=2$ of \eqref{hierarchy-LY} (though the case for $n=1$ is much simpler).
Note that the KdV hierarchy \eqref{hierarchy-LY} can be rephrased in the form
\ben
\frac{\pd}{\pd t_0}\left(\frac{1}{\lambda^{2}}\frac{2n+1}{2}\frac{\pd^2 F^c}{\pd t_0\pd t_n}
	- \frac{1}{8}\frac{\pd^4 F^c}{\pd t^3_0\pd t_{n-1}}
	- \frac{\pd^2 F^c}{\pd t^2_0}\frac{\pd^2 F^c}{\pd t_0\pd t_{n-1}} \right)
= -\frac{1}{2}\frac{\pd^3 F^c}{\pd t^3_0}\frac{\pd^2 F^c}{\pd t_0\pd t_{n-1}},
\een
whose $(\lambda^{-2})$-order terms of $\lambda$-expansion give
\bean
&\frac{\pd}{\pd t_0}
\left(\frac{2n+1}{2}\frac{\pd^2 F^c_{(1)}}{\pd t_0\pd t_n}
	- \frac{\pd^2 F^c_{(0)}}{\pd t^2_0}\frac{\pd^2 F^c_{(1)}}{\pd t_0\pd t_{n-1}}
	- \frac{\pd^2 F^c_{(1)}}{\pd t^2_0}\frac{\pd^2 F^c_{(0)}}{\pd t_0\pd t_{n-1}} 
	- \frac{1}{8}\frac{\pd^4 F^c_{(0)}}{\pd t^3_0\pd t_{n-1}} \right)
\\&= -\frac{1}{2}
\left(\frac{\pd^3 F^c_{(0)}}{\pd t^3_0}\frac{\pd^2 F^c_{(1)}}{\pd t_0\pd t_{n-1}}
+\frac{\pd^3 F^c_{(1)}}{\pd t^3_0}\frac{\pd^2 F^c_{(0)}}{\pd t_0\pd t_{n-1}}\right),
\enan
in particular for $n=2$
\be
\label{eq:lambda:2}
\frac{\pd}{\pd t_0}
\left(\frac{5}{2}\frac{\pd^2 F^c_{(1)}}{\pd t_0\pd t_2}
	- \frac{\pd^2 F^c_{(0)}}{\pd t^2_0}\frac{\pd^2 F^c_{(1)}}{\pd t_0\pd t_{1}}
	- \frac{\pd^2 F^c_{(1)}}{\pd t^2_0}\frac{\pd^2 F^c_{(0)}}{\pd t_0\pd t_{1}} 
	- \frac{1}{8}\frac{\pd^4 F^c_{(0)}}{\pd t^3_0\pd t_{1}} \right)
= -\frac{1}{2}
\left(\frac{\pd^3 F^c_{(0)}}{\pd t^3_0}\frac{\pd^2 F^c_{(1)}}{\pd t_0\pd t_{1}}
+\frac{\pd^3 F^c_{(1)}}{\pd t^3_0}\frac{\pd^2 F^c_{(0)}}{\pd t_0\pd t_{1}}\right),
\ee
with noting
\bean
	\frac{1}{a}\frac{\pd^2 F_{(1)}}{\pd t_0\pd t_2}
	&=\frac{P_3}{P_1^3}\frac{w^2}{2} - 2\frac{P_2^2}{P_1^4}\frac{w^2}{2}+2\frac{P_2}{P_1^3}w - \frac{1}{P_1^2},
	\\
	\frac{1}{a}\frac{\pd^2 F_{(1)}}{\pd t_0\pd t_1}
	&=\frac{P_3}{P_1^3}w - 2\frac{P_2^2}{P_1^4}w+2\frac{P_2}{P_1^3},
	\\
	\frac{1}{a}\frac{\pd^2 F_{(1)}}{\pd t_0^2}
	&=\frac{P_3}{P_1^3}-2\frac{P_2^2}{P_1^4},
	\\
	\frac{\pd^2 F_{(0)}}{\pd t_0 \pd t_{1}}
	&=\frac{w^2}{2},
\enan
where $P_n$ stands for $n$-th derivative of $P$ with respect to $w$. Then we have
\ben
\frac{1}{a}
\left(\frac{5}{2}\frac{\pd^2 F^c_{(1)}}{\pd t_0\pd t_2}
	- \frac{\pd^2 F^c_{(0)}}{\pd t^2_0}\frac{\pd^2 F^c_{(1)}}{\pd t_0\pd t_{1}}
	- \frac{\pd^2 F^c_{(1)}}{\pd t^2_0}\frac{\pd^2 F^c_{(0)}}{\pd t_0\pd t_{1}} \right)
= 
- \frac{1}{4}\frac{P_3}{P_1^3} w^2 
+ \frac{1}{2}\frac{P_2^2}{P_1^4} w^2
+3\frac{P_2}{P_1^3}w 
- \frac{5}{2}\frac{1}{P_1^2},
\een
giving the left hand side of \eqref{eq:lambda:2}:
\ben
\frac{1}{a}\frac{\pd}{\pd t_0}
\left(\frac{5}{2}\frac{\pd^2 F^c_{(1)}}{\pd t_0\pd t_2}
	- \frac{\pd^2 F^c_{(0)}}{\pd t^2_0}\frac{\pd^2 F^c_{(1)}}{\pd t_0\pd t_{1}}
	- \frac{\pd^2 F^c_{(1)}}{\pd t^2_0}\frac{\pd^2 F^c_{(0)}}{\pd t_0\pd t_{1}} \right)
=
 \frac{w^2}{4}\frac{P_4}{P_1^4} 
 - \frac{5}{2}\frac{P_3}{P_1^4} w
 +8\frac{P_2^2}{P_1^5}w 
 -8\frac{P_2}{P_1^4} 
-\frac{7}{4}\frac{P_2P_3}{P_1^5} w^2 
+2w^2\frac{P_2^3}{P_1^6} ,
\een
while the right hand side:
\ben
-\frac{1}{2a}
\left(\frac{\pd^3 F^c_{(0)}}{\pd t^3_0}\frac{\pd^2 F^c_{(1)}}{\pd t_0\pd t_{1}}
+\frac{\pd^3 F^c_{(1)}}{\pd t^3_0}\frac{\pd^2 F^c_{(0)}}{\pd t_0\pd t_{1}}\right)
=
\frac{w^2}{4}\frac{P_4}{P_1^4}
+\frac{1}{2}\frac{P_3}{P_1^4}w - \frac{P_2^2}{P_1^5}w + \frac{P_2}{P_1^4}-\frac{7w^2}{4}\frac{P_2P_3}{P_1^5}+2w^2\frac{P_2^3}{P_1^6},
\een
which gives us
\bean
&\frac{\pd}{\pd t_0}
\left(\frac{5}{2}\frac{\pd^2 F^c_{(1)}}{\pd t_0\pd t_2}
	- \frac{\pd^2 F^c_{(0)}}{\pd t^2_0}\frac{\pd^2 F^c_{(1)}}{\pd t_0\pd t_{1}}
	- \frac{\pd^2 F^c_{(1)}}{\pd t^2_0}\frac{\pd^2 F^c_{(0)}}{\pd t_0\pd t_{1}} \right)
	+\frac{1}{2}
\left(\frac{\pd^3 F^c_{(0)}}{\pd t^3_0}\frac{\pd^2 F^c_{(1)}}{\pd t_0\pd t_{1}}
+\frac{\pd^3 F^c_{(1)}}{\pd t^3_0}\frac{\pd^2 F^c_{(0)}}{\pd t_0\pd t_{1}}\right)
\\&
 =-3a\left(3\frac{P_2}{P_1^4}-3\frac{wP_2^2}{P_1^5}+\frac{wP_3}{P_1^4}\right),
\enan
whereas
\ben
	\frac{\pd^3 F_{(0)}}{\pd t_0^2 \pd t_{1}}=3\frac{P_2}{P_1^4}-3\frac{wP_2^2}{P_1^5}+\frac{wP_3}{P_1^4}.
\een
Therefore we finally obtain
\ben
\frac{1}{a}\frac{\pd}{\pd t_0}
\left(\frac{5}{2}\frac{\pd^2 F^c_{(1)}}{\pd t_0\pd t_2}
	- \frac{\pd^2 F^c_{(0)}}{\pd t^2_0}\frac{\pd^2 F^c_{(1)}}{\pd t_0\pd t_{1}}
	- \frac{\pd^2 F^c_{(1)}}{\pd t^2_0}\frac{\pd^2 F^c_{(0)}}{\pd t_0\pd t_{1}} \right)
	+\frac{1}{2a}
\left(\frac{\pd^3 F^c_{(0)}}{\pd t^3_0}\frac{\pd^2 F^c_{(1)}}{\pd t_0\pd t_{1}}
+\frac{\pd^3 F^c_{(1)}}{\pd t^3_0}\frac{\pd^2 F^c_{(0)}}{\pd t_0\pd t_{1}}\right)
=-3\frac{\pd^5 F_{(0)}}{\pd t_0^4 \pd t_{1}},
\een
by which the relation \eqref{eq:lambda:2} is reduced to
\ben
\frac{\pd}{\pd t_0}
\left( -3a\frac{\pd^4 F^c_{(0)}}{\pd t^3_0\pd t_{1}} 
	- \frac{1}{8}\frac{\pd^4 F^c_{(0)}}{\pd t^3_0\pd t_{1}} \right)
= 0,
\een
imposing $a=-\frac{1}{24}$.

\section*{Appendix C}

We describe computational details of \eqref{eq:proof:induc}: $\frac{\cL_n\tau}{\tau}+\mu_B\frac{\cL_{n-1}\tau}{\tau}=0\ (n=0,1,2)$.
Before going to showing it, we mention that from the usual Virasoro constraints \eqref{eq: Virasoro} it follows that
\bean
\frac{\cL_{-1}\tau}{\tau}
	&=-\frac{\d F^o}{\d t_0}+\sum_{n\ge 0}t_{n+1}\frac{\d F^o}{\d t_n}+\frac{\d F^o}{\d \mu_B}
	\quad\left(=0\quad\text{by open string equation}\right),
\\
\frac{\cL_0\tau}{\tau}
	&=-\frac{3}{2}\frac{\d F^o}{\d t_1}+\sum_{n\ge 0}\frac{2n+1}{2}t_n\frac{\d F^o}{\d t_n}
	-\mu_B\frac{\d F^o}{\d \mu_B}-\frac{1}{4},
\\
\frac{\cL_1\tau}{\tau}
	&=-\frac{5!!}{4}\frac{\d F^o}{\d t_{2}}+\sum_{i\ge 0}\frac{(2i+3)!!}{2^{2}(2i-1)!!}t_i\frac{\d F^o}{\d t_{i+1}}
	+\frac{u^2}{8}\left(\frac{\d^2 F^o}{\d t_0\d t_0}+\frac{\d F^o}{\d t_0}\frac{\d F^o}{\d t_0}+2\frac{\d F^o}{\d t_0}\frac{\d F^c}{\d t_0}\right)
	+\mu_B^{2}\frac{\d F^o}{\d \mu_B}+\frac{1}{2}\mu_B,
\\
\frac{\cL_2\tau}{\tau}
	&=-\frac{7!!}{8}\frac{\d F^o}{\d t_{3}}
	+\sum_{i\ge 0}\frac{(2i+5)!!}{2^{3}(2i-1)!!}t_i\frac{\d F^o}{\d t_{i+2}}\\
	&\qquad
	+u^2\frac{3!!}{8}\left(\frac{\d^2 F^o}{\d t_0\d t_1}+\frac{\d F^o}{\d t_0}\frac{\d F^o}{\d t_1}
		+\frac{\d F^o}{\d t_0}\frac{\d F^c}{\d t_1}+\frac{\d F^c}{\d t_0}\frac{\d F^o}{\d t_1}\right)
	-\mu_B^{3}\frac{\d F^o}{\d \mu_B}-\frac{3}{4}\mu_B^2.
\enan
Using the open KdV equations~\eqref{open-kdv-muB}
\ben
	\frac{2n+1}{2} F^o_n
	=-\mu_B F^o_{n-1}
	+\frac{\lambda^2}{2} F^o_0 F^c_{0,n-1}-\frac{\lambda^2}{4} F^c_{0,0,{n-1}},
\een
and the string equation \eqref{string-closed} with its derivatives
\ben
	F^c_{0}=\sum_{n\ge 0}t_{n+1}F^c_{n}+\frac{t_0^2}{2\lambda^2},
	\qquad
	F^c_{0,0}=\sum_{n\ge 0}t_{n+1}F^c_{0,n}+\frac{t_0}{\lambda^2},
	\qquad
	F^c_{0,0,0}=\sum_{n\ge 0}t_{n+1}F^c_{0,0,n}+\frac{1}{\lambda^2},
\een
with introducing a shorthand notation $F^c_{n}=\frac{\pd F^c}{\pd t_n}$, $F^c_{l,m,n}=\frac{\pd^3F^c}{\pd t_l\pd t_m\pd t_n}$ and so on,
for $n=0$ of \eqref{eq:proof:induc}, we get
\bean
\frac{\cL_0\tau}{\tau}+\mu_B\frac{\cL_{-1}\tau}{\tau}
	&=\left(-\frac{3}{2} F^o_1 -\mu_B F^o_0\right)
		+\frac{1}{2}t_0  F^o_0
		+\sum_{n\ge 1}t_{n}\left(\frac{2n+1}{2} F^o_n+\mu_B F^o_{n-1}\right)-\frac{1}{4}\\
	&=-\frac{\lambda^2}{2}F^o_0 F^c_{0,0}+\frac{\lambda^2}{4}F^c_{0,0,0}
		+\frac{1}{2}t_0F^o_0+\sum_{n\ge 1}t_n
		\left(\frac{\lambda^2}{2}F^o_0F^c_{0,{n-1}}-\frac{\lambda^2}{4}F^c_{0,0,{n-1}}\right)-\frac{1}{4}\\
	&=-\frac{\lambda^2}{2}F^o_0 \left(F^c_{0,0}-\sum_{n\ge 0}t_{n+1}F^c_{0,n}-\frac{t_0}{\lambda^2}\right)
		+\frac{\lambda^2}{4}\left(F^c_{0,0,0}-\sum_{n\ge 0}t_{n+1}F^c_{0,0,n}\right)-\frac{1}{4}
	=0.
\enan

Similarly, for $n=1$ of \eqref{eq:proof:induc},
with noting \eqref{open-constraint-muB}: $\mu_B=-\frac{\lambda^2}{2} \left(\left(  F^o_0\right)^2+ F^o_{0,0}+ 2F^c_{0,0}\right)$,
and the Virasoro constraint \eqref{eq: Virasoro} for $n=0$:
\ben
0=\frac{L_0 \tau^c}{\tau^c}=-\frac{3}{2} F^c_{1}+\sum_{n\ge 0}\frac{2n+1}{2}t_n  F^c_{n}+\frac{1}{16},
\een
where $\tau^c=\exp(F^c)$, yielding
\ben
	-\frac{3}{2} F^c_{0,1}+\sum_{n\ge 0}\frac{2n+1}{2}t_n  F^c_{0,n}
	=-\frac{1}{2}  F^c_{0},
	\qquad
	-\frac{3}{2} F^c_{0,0,1}+\sum_{n\ge 0}\frac{2n+1}{2}t_n  F^c_{0,0,n}
	=-  F^c_{0,0},
\een
we get
\bean
&\frac{\cL_1\tau}{\tau}+\mu_B\frac{\cL_{0}\tau}{\tau}\\
&=-\frac{3}{2}\left(\frac{5}{2} F^o_{2}+\mu_B F^o_1\right)
	+\sum_{n\ge 0}\frac{2n+1}{2}t_n\left(\frac{2n+3}{2} F^o_{n+1}+\mu_B F^o_n\right)
	+\frac{\lambda^2}{8}\left( F^o_{0,0}+\left(F^o_0\right)^2+2 F^o_0 F^c_0\right)
	+\frac{1}{4}\mu_B\\
&=-\frac{3}{2}\left(\frac{\lambda^2}{2} F^o_0 F^c_{0,1}-\frac{\lambda^2}{4} F^c_{0,0,1}\right)
	+\sum_{n\ge 0}\frac{2n+1}{2}t_n
	\left(\frac{\lambda^2}{2} F^o_0 F^c_{0,n}-\frac{\lambda^2}{4} F^c_{0,0,n}\right)\\
	&\qquad
	+\frac{\lambda^2}{8}\left( F^o_{0,0}+\left(F^o_0\right)^2+2 F^o_0 F^c_0\right)
	-\frac{\lambda^2}{8}\left(\left( F^o_0\right)^2+F^o_{0,0}+2 F^c_{0,0}\right)\\
&=\frac{\lambda^2}{2} F^o_0\left( -\frac{3}{2} F^c_{0,1}+\sum_{n\ge 0}\frac{2n+1}{2}t_n F^c_{0,n}\right)
	-\frac{\lambda^2}{4}\left(-\frac{3}{2} F^c_{0,0,1}+\sum_{n\ge 0}\frac{2n+1}{2}t_n F^c_{0,0,n}\right)
	+\frac{\lambda^2}{4} F^o_0 F^c_0-\frac{\lambda^2}{4} F^c_{0,0}\\
&=0.
\enan

Finally, for $n=2$ of \eqref{eq:proof:induc},
noting the Virasoro constraint \eqref{eq: Virasoro} for $n=1$: 
\ben
0=\frac{L_1 \tau^c}{\tau^c}
	=-\frac{5!!}{4} F^c_{2}+\sum_{n\ge 0}\frac{(2n+3)!!}{2^{2}(2n-1)!!}t_n F^c_{n+1}
	+\frac{\lambda^2}{8} F^c_{0,0}+\frac{\lambda^2}{8} \left(F^c_0\right)^2,
\een
and its derivatives, after some manipulations, we get
\ben
\frac{\cL_2\tau}{\tau}+\mu_B\frac{\cL_{1}\tau}{\tau}
=\frac{3\lambda^2}{8}F^c_{0,1}+\frac{\lambda^4}{16}F^c_{0,0}F^c_{0,0}+\frac{3\lambda^2}{8} F^o_1F^o_0
	+\frac{3\lambda^2}{16}F^o_{0,1}
	+\frac{\lambda^4}{16}F^o_{0,0}F^c_{0,0}
	-\frac{\mu_B^2}{4}
	+\lambda^2\frac{\mu_B}{8}F^o_0F^o_0.
\een
Furthermore, substituting
$\omega=F^c_{0,0}$, $\xi=F^o$, $F^c_{0,1}=\lambda^2\left(\frac{\omega^2}{2}+\frac{\omega_{2x}}{12}\right)$,
$\mu_B=-\frac{\lambda^2}{2}\left(\left(\xi_x\right)^2+\xi_{2x}+2\omega\right)$ 
where $\omega_{x}:=\omega_{0}$, $\omega_{2x}:=\omega_{0,0}$ and so forth,
and
\bean
\frac{3\lambda^2}{16}F^o_1
	&=\frac{\lambda^2}{8}\left(-\mu_B\xi_x+\frac{\lambda^2}{2}\xi_x\omega-\frac{\lambda^2}{4}\omega_x\right)
	=\frac{\lambda^4}{16}\left(\left(\xi_x\right)^3+\xi_x\xi_{2x}+3\xi_x\omega -\frac{\omega_x}{2} \right),
	\\
\frac{3\lambda^2}{16}F^o_{0,1}
	&=\frac{\lambda^2}{8}\left(-\mu_B\xi_{2x}+\frac{\lambda^2}{2}\xi_{2x}\omega+\frac{\lambda^2}{2}\xi_x\omega_x-\frac{\lambda^2}{4}\omega_{2x}\right)
	=\frac{\lambda^4}{16}\left( \left(\xi_x\right)^2\xi_{2x}+(\xi_{2x})^2+3\xi_{2x}\omega+\xi_x\omega_x-\frac{\omega_{2x}}{2}\right),
\enan
we find
\ben
\frac{1}{\lambda^4}\left(\frac{\cL_2\tau}{\tau}+\mu_B\frac{\cL_{1}\tau}{\tau}\right)
=-\frac{1}{\lambda^4}\frac{\mu_B^2}{4}
	+\frac{\omega^2}{4}
	+\frac{\left(\xi_x\right)^4}{16}
	+\frac{(\xi_x)^2\xi_{2x}}{8}
	+\frac{(\xi_x)^2\omega}{4}
	+\frac{(\xi_{2x})^2}{16}
	+\frac{\xi_{2x}\omega}{4}
=0.
\een


\end{document}